\documentstyle[aaspp4,11pt,tighten]{article}
\begin{document}
\input psfig.sty

\title{Resolving the Controversy Over the Core Radius of 47~Tucanae
(NGC~104)\footnote{Based on observations with the NASA/ESA Hubble Space
Telescope, obtained at the Space Telescope Science Institute, which is
operated by the Association of Universities for Research in Astronomy, Inc.,
under NASA contract NAS5-26555.}$^,$\footnote{Lick Observatory Bulletin
No.~139X.}}
\author{Justin H.\ Howell and Puragra Guhathakurta\footnote{Alfred P.\ Sloan
Research Fellow}}
\affil{UCO/Lick Observatory, Department of Astronomy \& Astrophysics,\\
University of California, Santa Cruz, California 95064, USA\\
Email: {\tt jhhowell@ucolick.org}, {\tt raja@ucolick.org}}
\and
\author{Ronald L.\ Gilliland}
\affil{Space Telescope Science Institute, 3700 San Martin Drive,\\
Baltimore, Maryland 21218, USA\\
Email: {\tt gillil@stsci.edu}}

\begin{abstract}
This paper investigates the discrepancy between recent measurements of the
density profile of the globular cluster 47~Tucanae that have used {\it Hubble
Space Telescope\/} data sets.  A large core radius would support the
long-held view that 47~Tuc is a relaxed cluster, while a small core radius
may indicate that it is in a post--core-collapse phase or possibly even on
the verge of core collapse, as suggested by a variety of unusual objects ---
millisecond pulsars, X-ray sources, high velocity stars --- observed in the
core of the cluster.  Guhathakurta et al. (1992) used pre-refurbishment Wide
Field/Planetary Camera~1 (WFPC1) $V$-band images to derive $r_{\rm
core}=23''\pm2''$.  \cite{calzetti}
suggested that the density profile is instead a superposition of two King
profiles, one with a small, $8''$ core radius and the other with a $25''$
core radius, based on $U$-band Faint Object Camera (FOC) images.  More
recently, \cite{guido} have used deep WFPC1 $U$-band images to
derive $r_{\rm core}=12''\pm2''$.  The cluster centers used in
these studies are in agreement with one another; differences
in the adopted centers are not the cause of the discrepancy.  Our independent
analysis of the data used by De~Marchi et~al.\ reaches the following
conclusions:
\begin{itemize}
\item[(1)]{De~Marchi et~al.'s $r_{\rm core}\sim12''$ value is
spuriously low, a result of radially-varying bias in the star counts in
a magnitude limited sample --- photometric
errors and a steeply rising stellar luminosity function cause more stars to
scatter across the limiting magnitude into the sample than out of it,
especially near the cluster center where crowding effects are most severe.}
\item[(2)]{Changing the limiting magnitude to the main sequence turnoff,
away from the steep part of the luminosity function, partially alleviates the
problem and results in $r_{\rm core}=18''$.}
\item[(3)]{Combining such a limiting magnitude with accurate photometry
derived from point-spread-function fitting, instead of the less accurate
aperture photometry employed by De~Marchi et~al., results in a
reliable measurement of the density profile which is well fit by $r_{\rm
core}=22''\pm2''$.}
\end{itemize}

\noindent
The Calzetti et~al.\ FOC-based density profile measurement is also likely to
have been biased by a poor choice of limiting magnitude and large radially
varying photometric errors associated with aperture photometry.
Archival Wide Field Planetary Camera~2 (WFPC2) data are used to derive a star
list with a higher degree of completeness, greater photometric accuracy, and
wider areal coverage than the WFPC1 and FOC data sets; the WFPC2-based
density profile supports the above conclusions, yielding $r_{\rm
core}=24\farcs0\pm1\farcs9$.
\end{abstract}


\section{Introduction}

Globular clusters are excellent laboratories for studying the dynamics of 
a stellar system.  The high density of stars near their centers results in
frequent interactions---e.g.,~single star-single star, single star-binary,
and binary-binary.  Such interactions redistribute energy throughout the
cluster and drive its global evolution on the so-called ``two-body
relaxation'' timescale.  This can sometimes be comparable to the typical
orbital period of stars near the cluster center, and significantly shorter
than the cluster age, implying that dynamical evolution is important.
The orbital period or crossing/dynamical time is related to the core radius
(the characteristic length scale associated with the inner density profile)
and the velocity dispersion of the cluster.  It is customary to characterize
the radial distribution of various stellar populations in terms of the core
radius (cf.~\cite{rasio}).

The inner stellar density profile of a globular cluster is suggestive
of its evolutionary state (\cite{hut96}).
Most clusters are characterized by constant
density cores, well fit by models based on a relaxed, Maxwellian distribution
function of stars out to the limiting (tidal) radius (\cite{king66a}).
On the other hand, about twenty percent of all globular clusters
appear to have undergone catastrophic gravothermal collapse
as a result of runaway energy loss from the core due to two-body relaxation
(\cite{dk86}; see the recent review in \cite{meyheg}).
Even these post--core-collapse (PCC) clusters, however, are {\it not\/}
expected to develop central singularities as heating by
binaries will drive a ``quasi-steady post-collapse'' stage followed by 
gravothermal oscillations.  Theoretical modeling suggests that clusters in
the post-collapse phase should have core radii in the range
$r_{\rm core}\sim0.01\>$--$\>0.04r_h$ (\cite{good87}; \cite{gao91}) or even
as large as $r_{\rm core}\sim0.09r_h$ (\cite{vc94}), where $r_h$ is the
half-mass radius.  These calculations are based on a single-mass stellar
population; models with a realistic stellar mass function could yield a
larger core radius.

The globular cluster 47~Tucanae has been studied extensively, and
has long been considered a prototypical relaxed globular cluster
(\cite{king85}) with a large core radius: $r_{\rm core}\sim25''$
(\cite{dk84}).  Recent observations of a variety of exotic objects, such as
millisecond pulsars (\cite{rob95}), X-ray sources (\cite{hg83};
\cite{hjv94}), and nine~stars whose velocities differ from the cluster mean
by $\gtrsim30$~km~${\rm s}^{-1}$ (\cite{karl95}),
have raised the possibility that 47~Tuc may be approaching
core collapse.  The half-light (or half-mass) radius of 47~Tuc is $r_h=174''$
(\cite{tdk93}), so if it has undergone core collapse it should have a core
radius in the range $1\farcs7$ to $16''$ (following \cite{good87}; 
\cite{gao91}; \cite{vc94}).  Two recent determinations of the core radius of
47~Tuc fall within the upper end of this range, as discussed in \S\,2 below.
Moreover, \cite{fisgeb95} have
constructed nonparametric dynamical models based on the surface brightness
and velocity dispersion profiles, and conclude that the mass profile of
47~Tuc is as centrally concentrated as that of M15.  The density
profile slope of 47~Tuc beyond $2'$ is found to be similar to that seen in
Fokker-Planck simulations of PCC clusters (\cite{cohn80}).

Ground-based studies typically use integrated surface brightness
measurements to determine a cluster's density profile (cf.~\cite{dk84}).  
A substantial
fraction of the optical light of a cluster comes from a handful of the
brightest red giant branch (RGB) stars
so that the effective Poisson error associated with
integrated light measurements is large.  This kind of ``sampling''
error makes the measured density profile noisy, and any error in the center
determination tends to bias the measured core radius towards large values.
{\it Hubble Space Telescope\/} ({\it HST\/}) images, even with the aberrated
pre-refurbishment point spread function (PSF), offer the advantage of
resolving individual stars down to the main sequence turnoff even in the
cluster core (Guhathakurta et~al. 1992, hereafter GYSB).  Star counts are
more representative of the stellar mass density than the integrated light,
and the effective Poisson errors are smaller (\cite{king66b}).  It is
preferable to work at short wavelengths
(e.g.,~the $U$ band) where the brightest RGB stars are suppressed relative
to the more numerous, bluer faint subgiants: this reduces sampling effects in
integrated light measurements and increases the degree of faint star
completeness (most notably in the vicinity of bright giants)
for star count measurements (Calzetti et~al.\ 1993; De Marchi et~al.\ 1996, 
hereafter DPSGB).

This paper focuses on the question: What is the core radius of 47~Tuc as
defined by the radial distribution of evolved stars?  It
examines recent measurements of the density profile of 47~Tuc 
that appear to be in disagreement with one another.  In particular, the {\it
HST\/} Wide Field/Planetary Camera~1 (WFPC1) data set analyzed by DPSGB
is reanalyzed here using somewhat different photometric
techniques; our results and those of DPSGB are compared to archival Wide
Field Planetary Camera~2 (WFPC2) data.  The background of the core radius
controversy is given in \S\,2.  In \S\,3, the available 47~Tuc data sets and
the methodologies of this paper and DPSGB are discussed.  In \S\,4 the
photometric errors associated with each method are examined, demonstrating
that the aperture photometry technique of DPSGB produces a radially-varying
bias in the star counts used to measure $r_{\rm core}$.
Core radius calculations are presented in \S\,5 along with tests which show
that the core radius discrepancy can be explained in terms of the star count
bias in the DPSGB study.  The conclusions of this paper are presented in \S\,6.

\section{The Controversy to Date}

The first {\it HST}-based core radius for 47~Tuc was derived from F555W
images obtained with the WFPC1 instrument operated in Planetary Camera
mode\footnote{All Wide Field/Planetary Camera~1 (WFPC1) data sets referred to
in this paper were obtained in Planetary Camera mode.}
(GYSB): $r_{\rm core}=23''\pm2''$, in
good agreement with earlier ground-based measurements (cf.~\cite{hr79}).
Shortly thereafter, \cite{calzetti} analyzed an independent set of
pre-refurbishment {\it HST\/} Faint Object Camera (FOC) ultraviolet data, and
obtained results that were in conflict with previous work.  They pointed out
that the cluster center determined from the ground was significantly in
error, biased by a few bright giant stars.  More surprisingly, they derived
a cluster density profile that did not resemble a King profile, but was
instead fit by a superposition of two King profiles with core radii of $25''$
and $8''$, suggesting that 47~Tuc is on the verge of core collapse.

The cause of the discrepancy between these density profile measurements
remained unclear for the next few years, and so did the true nature of
47~Tuc's density profile.  The center derived by \cite{calzetti} is $\sim6''$
from the earlier ground-based estimate (\cite{auriere}), but only $1\farcs4$
from GYSB's estimate which is well within the formal errors.
When Calzetti et~al.'s FOC ultraviolet image-based
cluster centroid estimate is used in conjunction with the GYSB F555W data
set, the resulting density profile is marginally smoother than the profile
obtained using GYSB's center estimate, possibly indicating that the former
center estimate is more accurate (see \S\,1), but the best-fit core radius is
$\gtrsim20''$ in both cases (\cite{gysb93}).  Two possible sources of
discrepancy in density profile measurements are ruled out:
\begin{itemize}
\item[(1)]{Detailed image simulations show that the sample of faint
RGB/subgiant stars used by GYSB is nearly complete and, in particular, the
degree of incompleteness does not vary appreciably with radius.}
\item[(2)]{If there were a bluer-inward radial gradient in the mean color
of faint RGB/subgiant stars, this would result in a difference between the
density profiles derived from ultraviolet versus visual-band images.
However, a star-by-star match between the Calzetti et~al.\ and GYSB data sets
reveals no such ($\rm F342W-F555W$) color gradient in 47~Tuc.}
\end{itemize}

More recently, DPSGB used an F336W image from the deep WFPC1 data set of
\cite{ron95} to revise the cluster center derived by \cite{calzetti} and
measure a core radius of $r_{\rm core}=12''\pm2''$, arguing that
neither the \cite{calzetti} composite King profile nor the traditional large
core radius fit the data.  This $12''$ core radius is within a factor of
two or three of the range of theoretical estimates for the core radius of a PCC
cluster with 47~Tuc's $r_h$.  
DPSGB argued that this is suggestive that 47~Tuc may be
approaching core collapse as \cite{calzetti} had proposed.

\section{Data and Photometric Techniques}

There are two {\it HST\/} imaging data sets analyzed in this paper:
({\bf 1})~pre-refurbishment WFPC1 data described by \cite{ron95},
consisting of a very deep F336W image ($\sim99\times1000$\,s) with excellent
sub-pixel dithering and shorter
F439W and F555W exposures (160\,s and 60\,s, respectively); and
({\bf 2})~archival WFPC2 data from program GO-6095 (PI: S.~G.~Djorgovski),
consisting of F218W ($4\times800$\,s), F439W ($2\times50$\,s), and F555W
(7\,s) images.  The deep WFPC1 data set combines the advantages of
ultraviolet imaging (as used by Calzetti et~al.\ 1993) with a wider field of
view (equivalent to that used by GYSB), and is significantly deeper than all
previous 47~Tuc
data sets.  The great improvement in PSF quality in the WFPC2 data
set relative to the deep WFPC1 data set more than compensates for the shorter
exposure times, resulting in a higher degree of completeness and about a
factor of two~smaller photometric errors (Guhathakurta et~al., in preparation).
Moreover, the larger
field of view of WFPC2 ($150''\times150''$ with the PC quadrant being
partially filled, extending to $r\sim100''$) compared to WFPC1
($68''\times68''$, extending to $r\sim55''$) makes it better for density
profile measurements.

Since this work and DPSGB use the same deep WFPC1 data set, it is important
to examine and compare the stellar photometry techniques used in the two
studies.
This paper uses standard PSF-fitting photometry techniques based on {\sc
daophot~ii} (\cite{stetson92}).  These techniques are described in detail in
GYSB and \cite{yanny94}, with minor modifications to adapt them to the
doubly-oversampled combined F336W WFPC1 image constructed from the sub-pixel
dithered exposures (Gilliland et~al.\ 1995; \cite{edmonds96}).
The main steps in the technique are
outlined here.  A preliminary star list is derived by applying
{\sc daophot}'s {\sc find}
routine to the deep WFPC1 F336W image.  For each of the four CCDs, a set of
bright, unsaturated, and relatively isolated stars are used to construct a
PSF template, which is allowed to vary quadratically with position.  This
template is then fit to all stars on the image using the preliminary star
list.  The neighbors of the PSF template stars are subtracted, and the
process is iterated a few times.  Stars too faint to be detected in the
initial {\sc find} run are then added based on inspection of the
PSF-subtracted frames.  The process of PSF template building and fitting is
iterated a few more times, and the final star list is fit.  The magnitudes
returned by {\sc daophot}'s PSF-fitting routine {\sc allstar} are converted
to total instrumental magnitudes by using the PSF template stars to determine a
spatially-dependent aperture correction.  The resulting photometry in the
F336W, F439W, and F555W bands is converted to the Johnson $UBV$ system
through an empirical fit to ground-based data from \cite{auriere94}.
A similar set of techniques is applied to the WFPC2 data set
(cf.~\cite{gysb96}) to derive an independent star list for
which the instrumental F218W, F439W, and F555W magnitudes are converted to
Johnson $U$, $B$, and $V$ magnitudes, respectively.

DPSGB uses a `core aperture photometry' technique defined in \cite{demarchi93}.
PSF peaks are identified on the deep WFPC1 F336W image and aperture
photometry is carried out using a small aperture comparable to the size of
the PSF core (hence the term `core aperture photometry')
and a sky annulus of roughly twice that radius. An aperture correction factor 
is applied to convert the resulting aperture magnitudes to total magnitudes,
compensating for the portion of the PSF that lies outside the aperture and
for the fact that the sky annulus includes light from the star in question.
This method is hereafter referred to as aperture photometry.
The sample of \cite{calzetti} is analyzed using
the same aperture photometry technique on the same crowded field.
While \S\,4 makes a specific comparison of our results to DPSGB because they
share a common data set, the conclusions are expected to apply to Calzetti
et~al.'s $r_{\rm core}$ analysis as well.  Neighbor contamination affects
aperture photometry to a much greater degree than it affects PSF-fitting
photometry.  Since aperture photometry is based on the total flux within
the aperture, the contribution of neighboring stars is a direct source of
photometric bias.  Neighbors have
a somewhat smaller effect on the sky measurement since the latter is
estimated from the mode.

There are three sets of stellar $U$ magnitudes discussed in this
study.  The photometry from this work using the deep WFPC1 data set is
referred to as $U\equiv{U_{\rm WFPC1}}$ (this paper); the
photometry presented by DPSGB based on the same data set is referred to as 
$U_{\rm D}\equiv{U_{\rm WFPC1}} ~({\rm DPSGB})-0.39$ (see \S\,\ref{sbscomp}), 
where $U_{\rm WFPC1} ({\rm DPSGB})$ is identical to the quantity $m_{336}$ in
that paper.
Photometry from the archival WFPC2 data set is referred to as $U_{\rm WFPC2}$.
The WFPC2 PSF is much sharper than that of WFPC1, resulting in more accurate
photometry; thus the WFPC2 data set is used to define the
`true' stellar magnitudes of the stars identified on both the WFPC1 and
WFPC2 images.  For each star in the WFPC1 and WFPC2 data sets, the positional
offset in arcseconds, $\Delta\alpha$ and $\Delta\delta$ (relative to
reference star `E' in the 47~Tuc core; GYSB), is determined using the IRAF
task {\sc metric} (\cite{gilmozzi}) which uses pointing information based on
the {\it HST\/} Guide Star Catalog.

\section{Photometric Error and its Effect on Star Counts}
\subsection{Star-by-Star Comparison}\label{sbscomp}

In order to compare the performance of the two photometric techniques described
above, the input star lists used in the two studies are examined.  If,
for example, one list systematically misses stars in the crowded 
central region of the cluster, then this will result in differences in
the measured core radii.
The star finding/PSF-fitting iterations described
above yield a set of 14801 stars identified
on the deep WFPC1 F336W image.  A star list kindly provided by
Guido De Marchi gives the positions and magnitudes of 14979 stars from
the same image as reduced by DPSGB.  The bright stars in these two star
lists match with an rms positional difference of one-third of one PC pixel,
$\sigma=0\farcs016$.  The entire matched star list contains 12372~stars with
positions agreeing to within $3\sigma$ (1~PC pixel)
between the two input star lists.  It should be noted that
141 stars appear twice on the star list from DPSGB.  The majority of the 
$\sim2500$
unmatched stars fall below the main sequence turnoff and thus are not
used to calculate $r_{\rm core}$.  An inspection
of the unmatched stars suggests that many such stars from
the DPSGB list may be spurious, as they tend to lie in the wings of bright
stars, while unmatched stars from this work tend to be faint stars
barely visible on the image.  The fraction of evolved stars (defined to be
those with $U<18.11$) that remain unmatched in each sample is quite small: 217
from this paper and 246 from DPSGB out of a total of $>4000$ evolved stars.
Comparison of our (DPSGB's) WFPC1 photometry to WFPC2 
photometry of these unmatched stars indicates that 10\% (40\%)
of the unmatched stars are main sequence stars scattered to brighter
magnitudes as a result of photometric error.  As expected, the
unmatched stars are such a small fraction of the total that they have a 
negligible effect on the $r_{\rm core}$ estimate
(\S\,5).  We conclude that the set of stars used in the $r_{\rm core}$
analysis is basically the same for our study and that of DPSGB; the sample
differences are too small to explain the vastly different conclusions about
$r_{\rm core}$.

Establishing a common photometric system is the next step in comparing the
photometry derived from 
the two data reduction techniques.  This paper converts instrumental magnitudes
to standard Johnson $UBV$ magnitudes empirically as described in \S\,3 above.  
DPSGB uses $m_{336}\equiv{U_{\rm WFPC1}}$~(DPSGB) throughout, referring to 
instrumental magnitudes
``converted to the WF/PC ground system''.  A star-by-star comparison between
the two data sets indicates that a constant offset of $-0.39$~mag applied to
$U_{\rm WFPC1}$ (DPSGB) brings the majority of stars into good agreement with
our measured $U$ magnitude.  The quantity $U_{\rm D}$ will hereafter be used
in place of $U_{\rm WFPC1}~ ({\rm DPSGB})-0.39$.

\begin{figure}
\plotfiddle{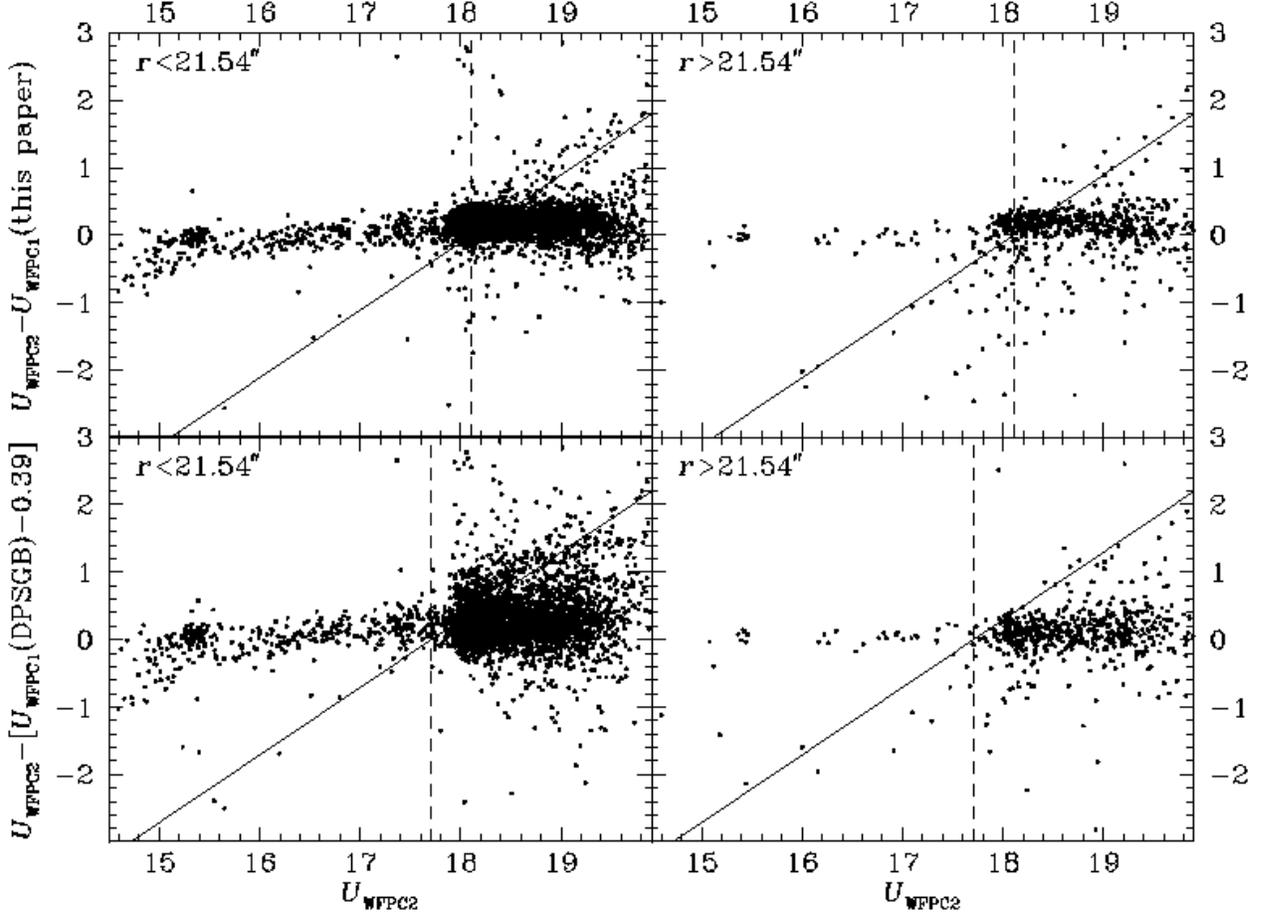}{4.0truein}{-90}{64}{64}{-243}{362}
\caption{Difference between WFPC1 and WFPC2 $U$-band stellar
photometry as a function of WFPC2 magnitude using the PSF-fitting method
(this work: upper panels) and the aperture photometry method (DPSGB:
lower panels).  The left panels show the inner sample of stars with
$r<21\farcs54$, while the right panels show the outer sample of stars with 
$r>21\farcs54$.  The smaller number of stars in the outer samples is due to
the limited overlap between the WFPC1 and WFPC2 fields of view away from the
cluster center.  The limiting magnitude used in core radius calculation is
indicated by the solid lines: $U^{\rm lim}=18.11$ (upper panels) and
$U_{\rm D}^{\rm lim}=17.71$ (lower panels).  Dashed lines show $U_{\rm
WFPC2}=18.11$ and $U_{\rm WFPC2}=17.71$ (upper and lower, respectively).  The
DPSGB aperture photometry displays a larger scatter than the PSF-fitting
photometry used in this work, and the scatter increases towards the cluster
center.  The mismatch between WFPC1 and WFPC2 photometry, most noticeable at
the bright end and near the turnoff, is a result of systematic error in
the WFPC1 to Johnson $U$ magnitude conversion.
\label{deltam}}
\end{figure}

To examine the variation of photometric scatter with radius, the WFPC1
magnitude ($U$ or $U_{\rm D}$) of each star is compared to its `true'
magnitude, $U_{\rm WFPC2}$.  The top and bottom panels of Fig.~\ref{deltam}
plot $(U_{\rm WFPC2}-U)$ and $(U_{\rm WFPC2}-U_{\rm D})$, respectively,
versus $U_{\rm WFPC2}$.
Only stars whose ($\Delta\alpha$, $\Delta\delta$) positions match to
within $0\farcs1$ between the WFPC1 and WFPC2 data sets are used.
The matched WFPC1 sample is divided at the median radius for
stars with $U<18.11$, $r_{\rm med}=21\farcs54$.  
The $r<r_{\rm med}$ (inner) subsamples are plotted in
the left panels and the $r>r_{\rm med}$ (outer)
subsamples are plotted in the right panels.  The small
number of points in the outer subsamples results from the limited overlap
between the WFPC1 and WFPC2 fields of view away from the core of the cluster.
There are systematic differences between the WFPC1 photometry and the WFPC2
photometry: $U$ and $U_{\rm D}$ are systematically too faint for bright red
giants and too bright for main sequence turnoff stars; these are caused by
errors in the conversion to Johnson $U$ magnitudes for the WFPC1 data
set (\S\,\ref{cmdsec}).

The  $U$ vs.\ $U_{\rm WFPC2}$ scatter is similar for inner and outer samples
while the scatter increases dramatically toward smaller radii for $U_{\rm D}$
vs.\ $U_{\rm WFPC2}$.  Moreover, DPSGB's photometry of turnoff stars has a
larger scatter than that of this work at all radii.  While the standard
deviation and mean value of $\Delta{U}$ $[\equiv(U_{\rm WFPC2}-U)$ or
$(U_{\rm WFPC2}-U_{\rm D})]$ are indicative of the overall rms photometric
scatter and
mean bias, respectively, it is instructive to examine the full distribution
of $\Delta{U}$ values.  Stars in the interval $17.8<U_{\rm WFPC2}<19.8$ are
used to construct cumulative distributions of $\Delta{U}$.  The cumulative
distributions using DPSGB's photometry 
show a long, asymmetric tail of large $(U_{\rm WFPC2}-U_{\rm D})$, 
while the distributions using the photometry from this study are quite
symmetrical and narrow, implying lower scatter and little or no bias.  
More quantitatively, the $\Delta{U}$ values corresponding to
the 95th percentile of the distribution are: +0.5, +0.5 (this study: inner
and outer subsamples, respectively); +1.2, +0.75 (DPSGB: inner and outer
subsamples, respectively).

The conclusion to be drawn from this subsection is that the star lists
used in the two studies are effectively identical, but
photometric accuracy is not.  DPSGB's photometry shows a systematically
greater photometric error, and a bias in the sense that stars are more
likely to be measured as brighter than they truly are than fainter.  More
importantly, this error and bias are greatest near the cluster center.

\subsection{Color-Magnitude Diagrams}\label{cmdsec}

\begin{figure}
\plotone{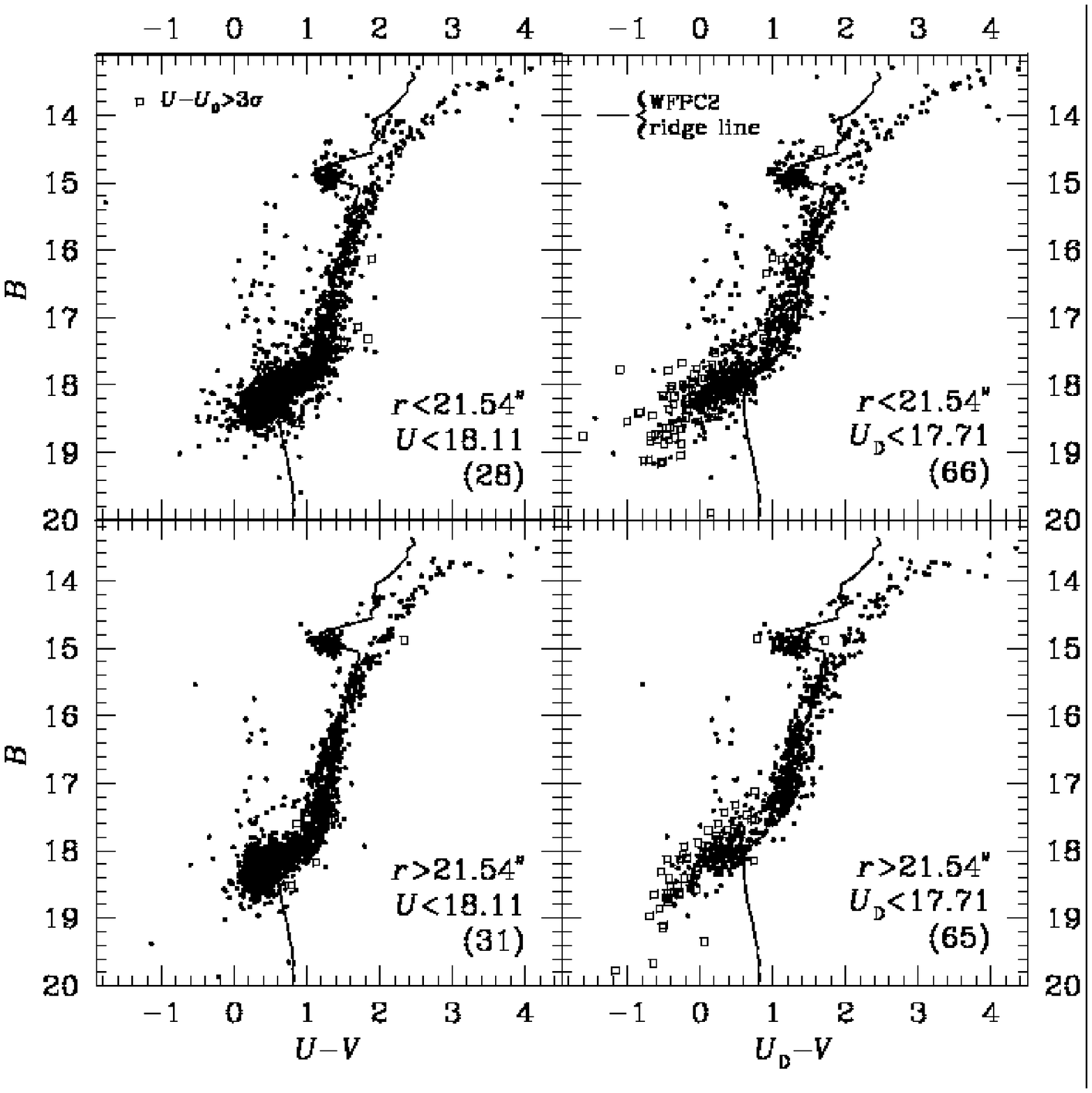}
\caption{Color-magnitude diagrams for the inner (top panels)
and outer (bottom panels) WFPC1 subsamples, using $U$-band photometry from
this work ($U$: left) and DPSGB ($U_{\rm D}$: right), combined with $B$- and
$V$-band photometry derived from PSF-fitting.  The limiting magnitude is
$U^{\rm lim}=18.11$ for this work, and $U_{\rm D}^{\rm lim}=17.71$ for DPSGB.
Squares represent stars which are $>+3\sigma$ outliers in a $U-U_{\rm D}$
scatter plot (their number is indicated in parentheses in each panel).  These
outliers tend to have bluer $U_{\rm D}-V$ colors than the main CMD features,
but their $U-V$ colors are not similarly biased, indicating that they are
outliers because of photometric scatter/bias in $U_{\rm D}$.  The solid line
is the median color of stars derived from WFPC2 photometry.  The disagreement
between WFPC2 photometry and both sets of WFPC1 photometry, most pronounced
for bright red giants and turnoff stars, is caused by systematic error in
the WFPC1 to Johnson $U$ calibration.
\label{cmd}}
\end{figure}

Color-magnitude diagrams (CMDs; Fig.~\ref{cmd}) provide an alternate
perspective on the nature of photometric error in the WFPC1 $U$-band data
set, complementary to the discussion in the previous subsection.  As
described below, the CMDs also help to highlight a discrepancy between the
WFPC1 and WFPC2 $U$-band photometric calibration.
The CMDs combine $B$ and $V$ magnitudes, obtained via PSF-fitting,
with each of the WFPC1 $U$ magnitudes under examination: our PSF-fitting
photometry and DPSGB's aperture photometry.  The $B$ vs.\ $U-V$ and
$B$ vs.\ $U_{\rm D}-V$ CMDs (left and right, respectively) are presented
for stars in the inner and
outer halves of the matched star sample (upper and lower, respectively).
The $B$ vs.\ $U-V$ diagrams use $U^{\rm lim}=18.11$, 
while the $B$ vs.\ $U_{\rm D}-V$ diagrams use
$U_{\rm D}^{\rm lim}=17.71$.  Limiting magnitudes will be discussed in 
detail in \S\,\ref{lfsec}.

The use of the CMD as a diagnostic tool is best
illustrated through stars with the most extreme photometric errors.
A plot of $U-U_{\rm D}$ vs.\ $U_{\rm WFPC2}$ is
used to identify `outliers' with
$U-U_{\rm D}>+3\sigma$, where $\sigma$ is calculated in running bins of width
0.5~mag in $U_{\rm WFPC2}$ (open squares in Fig.~\ref{cmd}); the $U-U_{\rm
D}$ distribution is skewed towards positive values such that very few stars
have $U-U_{\rm D}<-3\sigma$.
The open squares in $B$ vs.\ $U_{\rm D}-V$ diagrams generally lie on or
beyond the blue fringe of the subgiant branch and RGB defined by the rest of
the points.  By contrast, the vast majority of the open squares in the $B$
vs.\ $U-V$ diagrams are distributed evenly about the underlying subgiant and
red giant branches, with only a few points near the red edge.  This indicates
that outliers in the $U-U_{\rm D}$ distribution are predominantly the result
of the measured $U_{\rm D}$ value being systematically too bright.  As
expected, the greater degree of crowding causes outliers in the inner
subsample to display larger deviations in the CMD
than those in the outer subsample.  Moreover, the width of
the RGB appears to be slightly greater in $U_{\rm D}-V$ than in $U-V$.

The solid lines show the median color of stars from the
WFPC2 sample in 0.1~mag bins.  The discrepancy between WFPC2 and
WFPC1 photometry for both this study and DPSGB is worst for the reddest stars
(bright RGB) and for the bluest stars (turnoff), and in opposite senses. 
This results from error in the Johnson $U$ calibration for the WFPC1 data set,
which has been improved for the WFPC2 data set.
Both data sets have been empirically
transformed to the Johnson system using the ground-based data of 
Auri\`{e}re et~al. (1994).  However, the larger field of view of WFPC2 includes 
stars farther from the cluster center and thus less affected by crowding in 
Auri\`{e}re et al.'s data.  The WFPC2 ridge line is in good agreement with
Auri\`{e}re et al.'s photometry, which should be accurate outside the crowded 
cluster core.

\subsection{Luminosity Function}\label{lfsec}

\begin{figure}
\plotfiddle{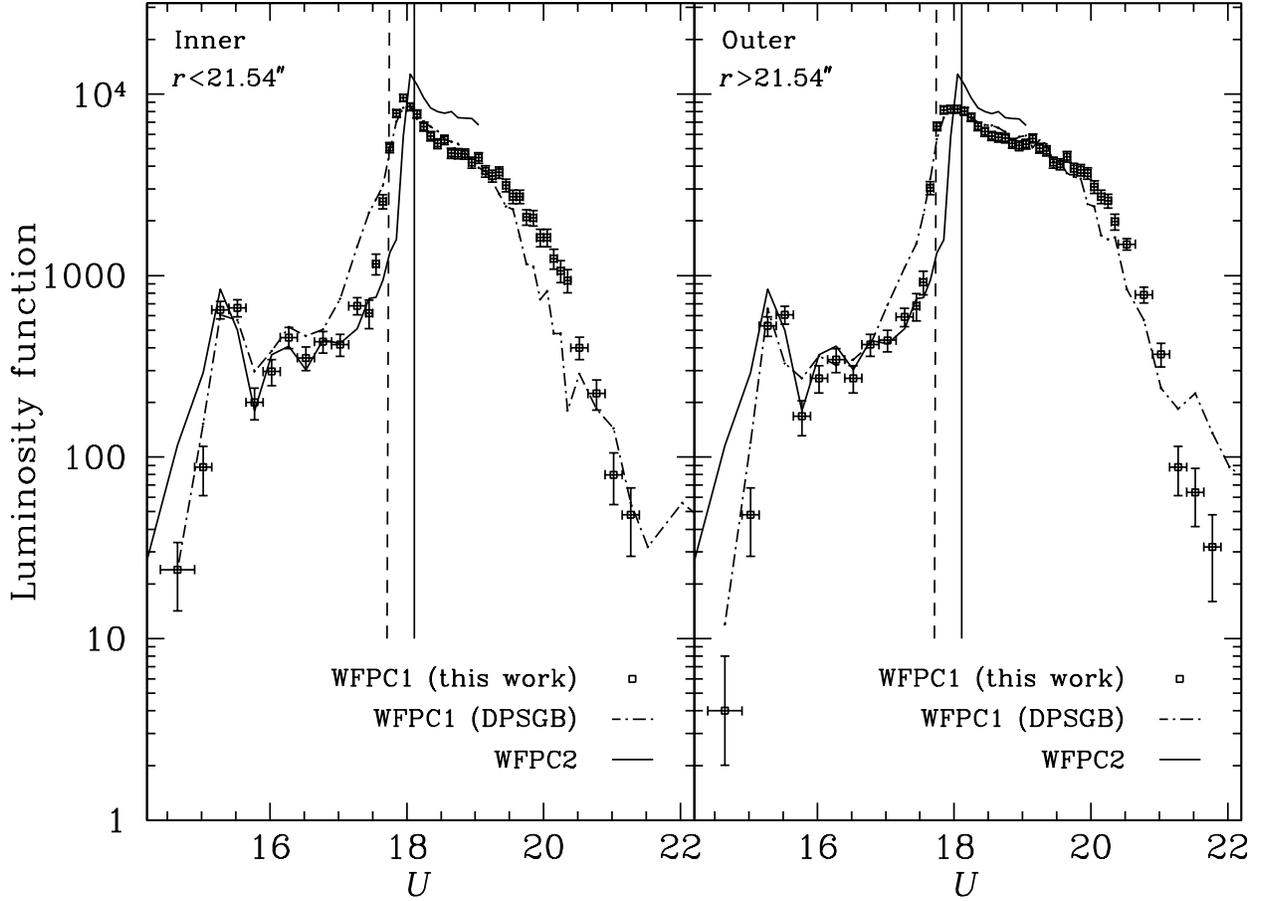}{4.0truein}{-90}{64}{64}{-243}{402}
\caption{Stellar $U$-band luminosity functions for the inner
(left) and outer (right) subsamples.  Points with error bars are based on
PSF-fitting photometry derived from a deep WFPC1 image (this work), the
dot-dashed line is based on aperture photometry on the same deep WFPC1
data set (DPSGB), and the bold solid line is derived from archival WFPC2
data.  The WFPC2 luminosity function is the same in both panels, with
different empirical normalizations to match the inner and outer WFPC1
luminosity functions.  The solid vertical line at $U=18.11$ and the
dashed line at $U=17.71$ mark the limiting magnitudes used in this work
and by DPSGB, respectively.
\label{lf}}
\end{figure}

In this subsection, we examine the stellar luminosity function of 47~Tuc.
This is a key element in understanding the bias in star counts in a
magnitude-limited sample that results from photometric error (scatter and
bias), and therefore in understanding the difference in core radius
estimates between DPSGB and GYSB.
The $U$-band stellar luminosity functions for the inner and outer samples 
are shown in the left and right panels of Fig.~\ref{lf}, respectively.
Data points represent the full $U$-band sample from this paper, the 
dot-dashed line
is the DPSGB $U_{\rm D}$ sample excluding the 141 duplicate stars, and the
solid line is the WFPC2 sample for $U_{\rm WFPC2}\le19$, normalized to match
the WFPC1 luminosity functions.  The WFPC2 luminosity functions shown in the
two panels are identical
except for the normalization, and are based on the full star list derived
from the entire WFPC2 field of view.  Since this study relies only on
post--main-sequence stars which are {\it not\/} expected to be affected by
mass segregation, we have chosen to combine the entire WFPC2 sample into a
single luminosity function; a full study of mass segregation will be
presented in Guhathakurta et~al. (in preparation).

In order to isolate a sample of post--main-sequence stars, the $U$ magnitude
of the main sequence turnoff is determined using a sample of turnoff stars
identified in a WFPC2 $V$ vs.\ $B-V$ CMD.  The mean $U_{\rm WFPC2}$ magnitude
of these stars is adopted as the limiting magnitude for the deep WFPC1 data
set analyzed in this paper: $U^{\rm lim}=18.11$.  On the other hand, DPSGB
used $m_{336}^{\rm lim}\equiv{U}_{\rm WFPC1}^{\rm lim}~ ({\rm DPSGB})=18.1$
which corresponds to $U_{\rm D}^{\rm lim}=17.71$.
Vertical lines are drawn at $U=18.11$ and $U=17.71$ in  Fig.~\ref{lf}; these
correspond to the diagonal lines in Fig.~\ref{deltam}.  Note that
the magnitude cut used by DSPGB falls on a steep part of the luminosity
function.  The next subsection quantifies the consequences of choosing a
limiting magnitude on a steep part of the luminosity function.

Of the three samples, WFPC2 photometry produces the sharpest
luminosity function.  The luminosity function derived from this study's
WFPC1 photometry has a somewhat more gradual rise from the RGB to the main
sequence turnoff than the WFPC2 luminosity
function as a result of its larger photometric errors.  The DPSGB luminosity
functions, particularly their inner sample, have the most gradual transition
between the RGB and turnoff, indicating the largest photometric errors.
There are significant differences between the
WFPC2 luminosity function and those derived from the deep WFPC1 data (our
study and DPSGB) most notably at the bright end and near the turnoff; these
are a result of systematic error in the conversion to Johnson $U$ for the
WFPC1 data set (\S\,\ref{cmdsec}).

\subsection{Discussion of Errors and ``Limiting Magnitude Bias''}

The above diagnostics (\S\S\,4.1--4.3) complement one another by
providing somewhat different perspectives on the photometric error in the
various data sets.  The $\Delta{U}$ plots indicate the photometric scatter as 
a function of
radius for each of the deep WFPC1 photometry sets, but only under the
{\it assumption\/} that $U_{\rm WFPC2}$ is the true magnitude of the
star.  While it is possible in principle to
compare the three data sets to empirically
determine the photometric error in each, the strongly non-gaussian error
distributions make this impractical.
Instead, the CMD diagnostic described in \S\,4.2 clearly
demonstrates that there are large photometric errors associated with the
$U_{\rm D}$ photometry but not with $U$, and this conclusion is independent
of the accuracy of $U_{\rm WFPC2}$.

A skeptic may wonder if there are systematic errors generic to the PSF-fitting
method that `cancel' each other in the $B$ vs.\ $U-V$ CMD or in the
$U$ vs.\ $U_{\rm WFPC2}$ comparison causing the PSF-fitting magnitudes to
appear more accurate than they actually are.  A related question is:  How
accurate is $U_{\rm WFPC2}$?~~  The luminosity function diagnostic is
useful in this regard, despite the mismatches caused by errors in the WFPC1
Johnson $U$ conversion.  Figure~\ref{lf} clearly shows that the most
prominent features (e.g.,~the steep rise at the subgiant branch, the peak
at the main sequence turnoff) are sharpest for WFPC2 photometry, slightly
smoothed out by photometric errors for our deep WFPC1 photometry, and
even more smoothed out (larger photometric errors) in the case of the
DPSGB data.

Photometric errors associated with the aperture photometry technique result
in a significant, radially-varying bias in the star counts; we hereafter
refer to this as ``limiting magnitude bias''.  The limiting magnitude used by
DPSGB is on a steep part of the luminosity function where many
stars which are slightly fainter than the limiting magnitude scatter into the
sample while relatively few stars scatter out of the sample.  Limiting
magnitude bias is exacerbated by the asymmetry in the distribution of
photometric errors (photometric bias; see \S\,4.1) and by the fact that the
errors (scatter and bias) tend to increase towards the fainter magnitudes.
Since aperture photometry errors increase with stellar crowding towards the
cluster center, so does the degree of limiting magnitude bias.

The differential limiting magnitude bias between inner and outer
subsamples may be quantified as follows.  Since the bounding radius of the
inner sample, $r_{\rm med}$, was chosen to be the median radius of stars in the
matched sample with $U<18.11$, the inner to outer star count ratio is
expected to be unity.  It is not surprising that $N(r<r_{\rm med})/N(r>r_{\rm
med})=1.02\pm0.03$ for stars from this study with $U<18.11$.  By contrast,
stars with $U_{\rm D}<17.71$ from the DPSGB sample have $N(r<r_{\rm
med})/N(r>r_{\rm med})=1.42\pm0.06$.  The latter ratio is significantly
greater than unity, indicating radially-varying limiting magnitude bias in
the DPSGB sample.

Alternatively, the limiting magnitude bias for each WFPC1 sample can be
quantified with respect to the `true' WFPC2 photometry.  The diagonal lines
in Fig.~\ref{deltam} correspond to $U=U^{\rm lim}$ (upper panels) and
$U_{\rm D}=U_{\rm D}^{\rm lim}$ (lower panels).  The vertical dashed lines
indicate $U_{\rm WFPC2}=U^{\rm lim}$ (upper panels) and $U_{\rm WFPC2}=U_{\rm
D}^{\rm lim}$ (lower panels).  The ratios $N(U<U^{\rm lim})/N(U_{\rm
WFPC2}<U^{\rm lim})$ are $1.54\pm0.06$ and $1.34\pm0.15$ for inner and outer
subsamples, respectively, while $N(U_{\rm D}<U_{\rm D}^{\rm lim})/N(U_{\rm
WFPC2}<U_{\rm D}^{\rm lim})$ are $2.32\pm0.14$ and $1.09\pm0.22$ for inner
and outer subsamples, respectively.  These ratios are greater than unity
because errors in the WFPC1 Johnson $U$ conversion cause the main sequence
magnitudes to be systematically too bright (\S\,\ref{cmdsec};
Fig.~\ref{deltam}).  Each ratio indicates the number of stars selected in
a sample relative to the corresponding true (WFPC2-based) number of stars
satisfying the selection criterion.  For the photometry from this work, the
excess relative to unity is similar in the inner and outer subsamples, so the
derived core radius should not be greatly affected.  DPSGB's photometry
results in a large and significant excess of stars selected in the inner
subsample, but not in the outer subsample.  This radial variation has a
substantial impact on the core radius measurement (\S\,\ref{corecalc}).
The large difference between the inner and outer DPSGB ratios is a direct
result of the increased scatter and bias in their photometry at small radii.

\section{Core Radius Calculations}\label{corecalc}

The effect of limiting magnitude bias on the measurement of the core radius
is explored in this section.  A one-parameter surface density profile of the
form:
\begin{equation}
\sigma{(r)}={{\sigma_0}\over{[1+(r/r_{\rm core})^2]}}
\end{equation}
is used to make maximum likelihood fits to various magnitude-limited samples
of evolved stars to determine $r_{\rm core}$, while the normalization
constant $\sigma_0$ is constrained by the total number of stars in each
sample.  Calculations are performed using both GYSB's estimate of the cluster
center:
\[
\alpha{_{\rm J2000}}=00^{\rm h}24^{\rm m}05^{\rm s}\!\!.87
\]
\begin{equation}
\delta{_{\rm J2000}}=-72^{\circ}04^{\prime}57\farcs8
\end{equation}
and DPSGB's center estimate:
\[
\alpha{_{\rm J2000}}=00^{\rm h}24^{\rm m}05^{\rm s}\!\!.29
\]
\begin{equation}
 ~~~~~\delta{_{\rm J2000}}=-72^{\circ}04^{\prime}56\farcs3~~~~.
\end{equation}
The two center estimates agree to within the quoted uncertainty of
$\sim1\farcs0$--$1\farcs3$ in each estimate.  Unless otherwise stated, all
core radius estimates in this paper are based on the GYSB center.

Table~\ref{table1} contains a summary of the core radius measurements.
The three letters identifying each calculation indicate the source of:
(1)~the star list (D for DPSGB, H for this work, M for the matched sample);
(2)~the photometry (D or H); and (3)~the limiting magnitude (D or H).
For direct comparison with DPSGB's $r_{\rm core}$ measurement, the 141
duplicate stars are included in their star list.  The result of the maximum
likelihood fit (case DDD) is $r_{\rm core}=13\farcs7\pm1\farcs8$,
in agreement with the DPSGB value of $r_{\rm core}=12\farcs2\pm2\farcs1$ 
derived using radial binning and a least-squares fit.  
A maximum likelihood fit to
the evolved star sample from this paper (case HHH) gives 
$r_{\rm core}=21\farcs8\pm2\farcs0$.
Applying a limiting magnitude of $U_{\rm D}^{\rm lim}=18.11$ instead of
$U_{\rm D}^{\rm lim}=17.71$ to DPSGB's star list and photometry
results in a core radius
of $17\farcs4\pm1\farcs5$ (case DDH).  The change in derived core radius
from $13\farcs7$ to $17\farcs4$
is a direct result of the change in
limiting magnitude and is independent of whether
the 141 duplicate stars are included.
Using the star list and photometry from this study but choosing DPSGB's
limiting magnitude $U^{\rm lim}=17.71$
gives $r_{\rm core}=21\farcs0\pm3\farcs3$ (case HHD).
Thus, the large scatter/bias in DPSGB's photometry and their increase
towards the crowded cluster center, combined with a poor choice of limiting
magnitude, are responsible for the spuriously low core radius estimate.

Table~\ref{table1} also shows that the measured $r_{\rm core}$ is independent
of whether the star list from this study, DPSGB's star list, or the matched
star list is used.  The dependence of the derived $r_{\rm core}$ on limiting
magnitude and photometric technique is roughly the same for all three lists.
Likewise the choice of cluster center has little effect on the
derived $r_{\rm core}$: although core radii based on the DPSGB center tend to
be $\approx0\farcs5$ smaller than those based on GYSB's center, this
difference is less than the uncertainty in an individual measurement.

\begin{figure}
\vskip -2.0truein
\centerline{\psfig{figure=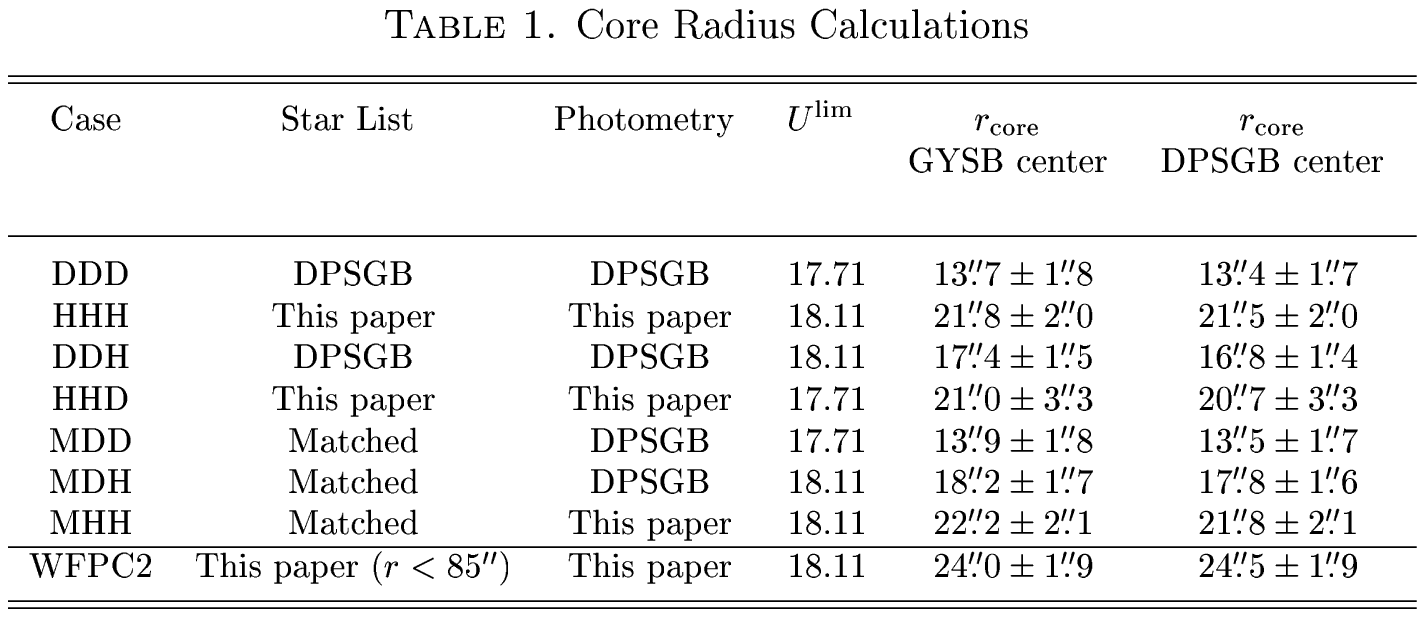}}
\end{figure}
\begin{table}\dummytable\label{table1}\end{table}

\begin{figure}
\plotfiddle{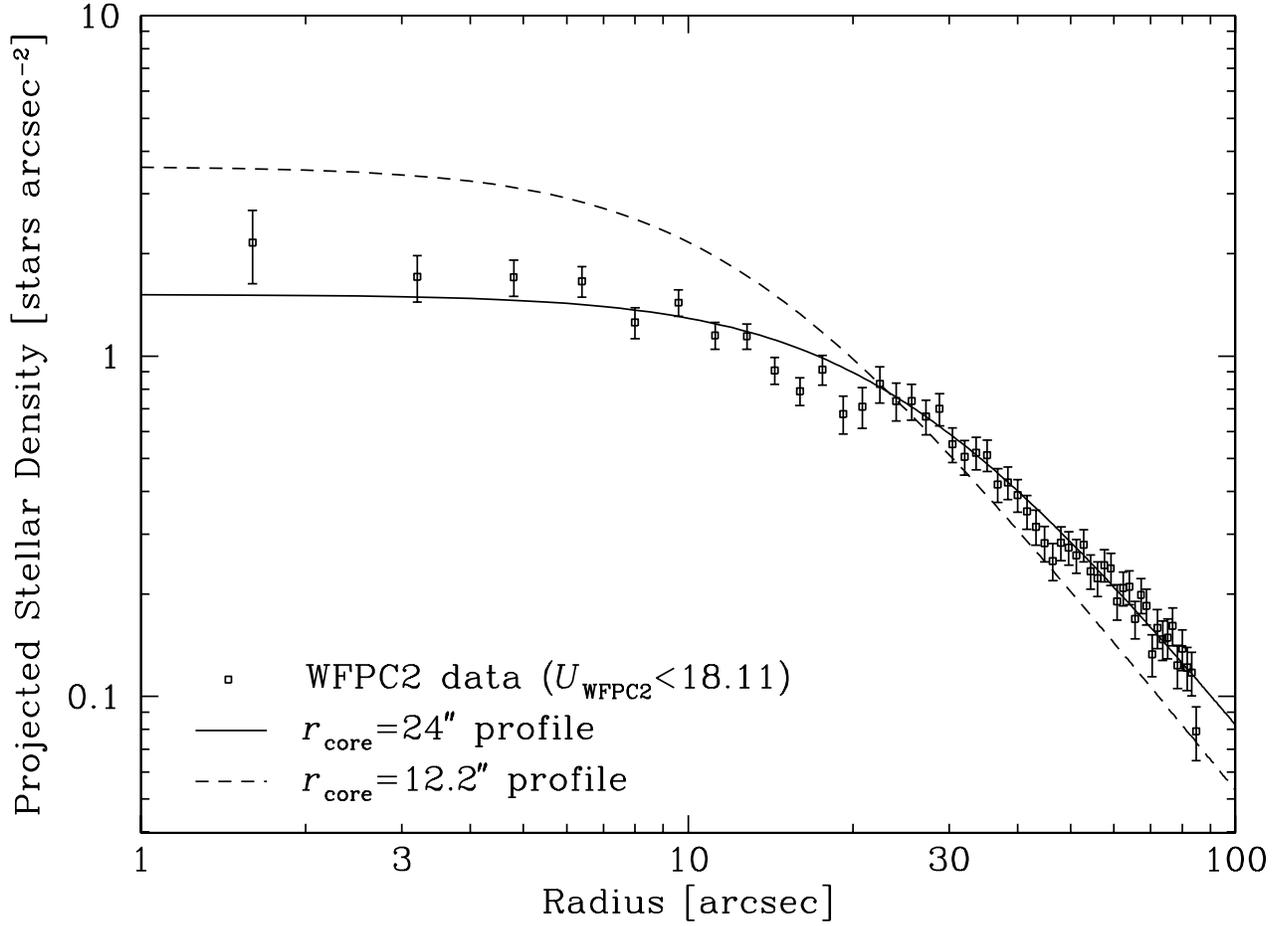}{4.0truein}{-90}{64}{64}{-243}{402}
\caption{Radial surface density profile based on stars
derived from the archival WFPC2 data set (squares with $1\sigma$ Poisson
error bars).  The solid and dashed curves represent profiles with core
radii of $24''$ and $12\farcs2$, respectively, each normalized to the
observed number of stars.  The former profile is a good fit to the data,
while the latter is clearly inconsistent with the data.
\label{wfpc2fig}}
\end{figure}

\begin{figure}
\plotfiddle{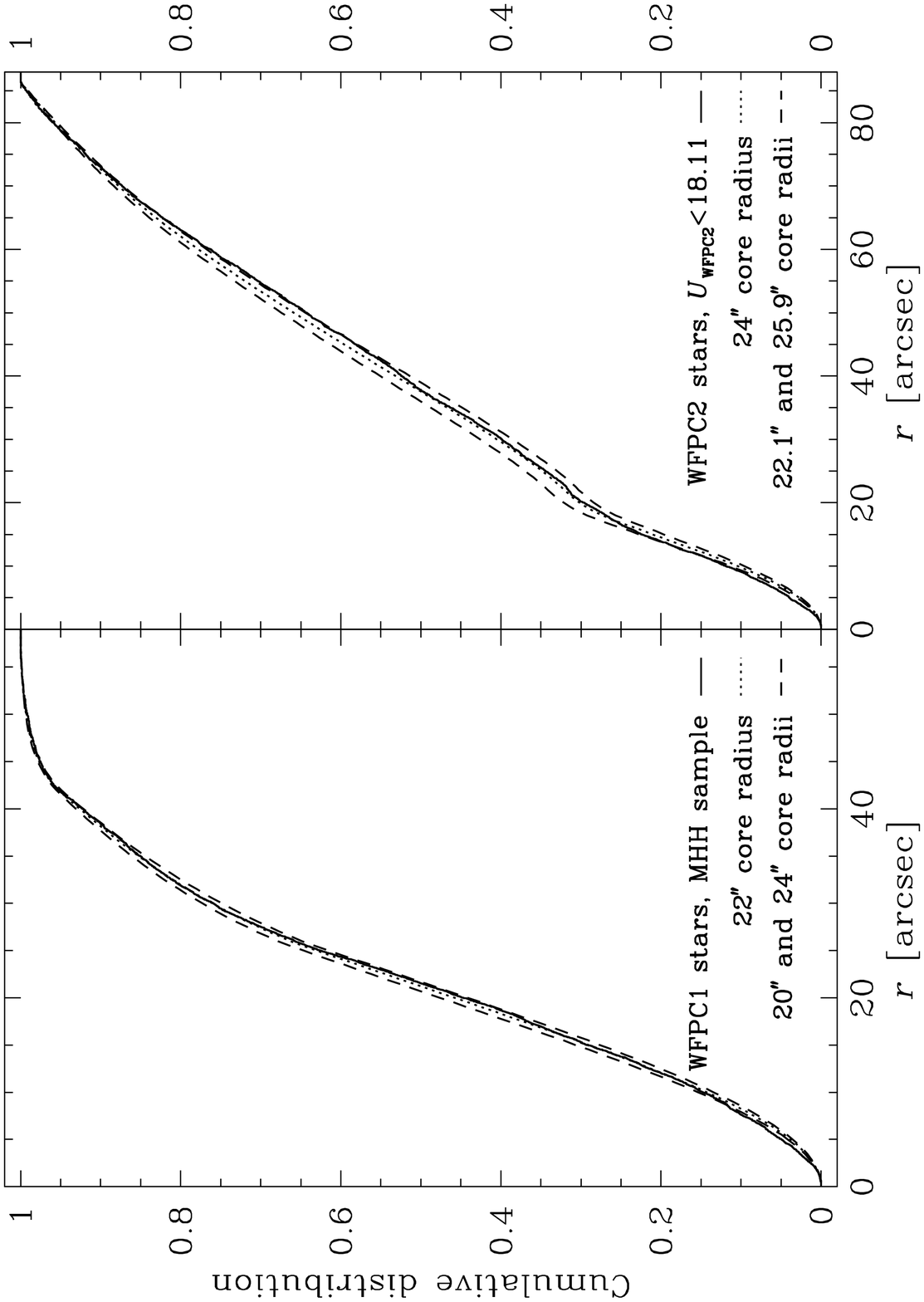}{4.0truein}{-90}{64}{64}{-243}{402}
\caption{Cumulative radial distribution of the matched deep
WFPC1 and truncated WFPC2 samples (solid lines, left and right panels,
respectively).  Best fit model profiles with $r_{\rm core}=22''$ and
$r_{\rm core}=24''$ are indicated by dotted lines in the left and right
panels, respectively; the dashed lines indicate model profiles with $r_{\rm
core}$ values that are $\pm1\sigma$ removed from the best fit value.  The
shapes of the WFPC1 and WFPC2 cumulative distributions are different because
of the difference in geometry between the two data sets and because of
the $r\sim85''$ truncation of the WFPC2 data set.
\label{cumdist}}
\end{figure}

The WFPC2 data set provides a consistency check on the core 
radius of 47~Tuc.  For all stars derived from the archival WFPC2 data set with
$U_{\rm WFPC2}<18.11$, the same maximum likelihood algorithm used above
yields $r_{\rm core}=23\farcs1\pm1\farcs7$.
An independent sample of 7044~turnoff stars with
$17\lesssim{V}<18$ drawn from the
WFPC2 $V$ vs.\ $B-V$ CMD (avoiding potential blend artifacts that may have
scattered to brighter magnitudes) results in $r_{\rm
core}=23\farcs3\pm1\farcs2$.  While these turnoff stars are fainter on
average than the evolved stars used in earlier density profile studies, they
are expected to have roughly the same mass as red giants so that mass
segregation effects are unlikely to be important.  A third calculation
truncates the $U_{\rm WFPC2}<18.11$ sample at $r\sim85''$ as a precaution
against inter-CCD edge effects caused by vignetting near the ridges of the
pyramid mirror.  A core radius of $r_{\rm core}=24\farcs0\pm1\farcs9$ (case
WFPC2 in Table~\ref{table1}) is derived from this truncated sample.
It is reassuring that these WFPC2 $r_{\rm core}$ measurements are
consistent with GYSB, earlier ground-based measurements, and the HHH and MHH
calculations above.  Figure~\ref{wfpc2fig} shows the radial surface
density profile of the WFPC2 truncated sample.
Also shown are an $r_{\rm core}=24\farcs0$ profile
(solid line) and an $r_{\rm core}=12\farcs2$ profile (dashed line); the 
latter is clearly inconsistent with the data.
These profiles have been normalized to the total observed number of stars
in the plot.  Figure~\ref{cumdist} shows the cumulative radial distributions
of the MHH sample (left panel) and the truncated WFPC2 sample (right panel).
Also shown are profiles for the best fitting core radii (dotted lines) and 
for core radii that differ from the best fit value by $\pm1\sigma$ (dashed
lines).

As an alternative to the maximum likelihood technique, we conduct
Kolmogorov-Smirnov tests between the star count samples and the fitting
function in Eq.~(1).  The MHH sample of 4504~stars (Table~\ref{table1}) and
the truncated WFPC2 sample of 3963~stars
are used to construct cumulative radial distributions
(Fig.~\ref{cumdist}).  One-sided Kolmogorov-Smirnov tests
indicate that the MHH data differ from the $r_{\rm core}=22''$ fitting
function by an amount that would be exceeded by chance 26\% of the time.
Similarly, the truncated WFPC2 data differ from an $r_{\rm core}=24''$
profile by an amount that would be exceeded by chance 33\% of the time.
Thus, the functional form adopted here provides an adequate fit to the star
count data.

As a final check on the $r_{\rm core}$ calculations, one can test
whether the degree of limiting magnitude bias (\S\,4) for a given sample is
sufficient to explain the core radius derived for it.  Using the 
photometry of DPSGB and their sample of stars with $U_{\rm D}<U_{\rm D}^{\rm
lim}$ (case DDD in Table~\ref{table1}), the inner to outer star count ratio
is $N(r<r_{\rm med})/N(r>r_{\rm med})=1.42\pm0.06$.  Integrating a $14''$
profile (comparable to the maximum likelihood fit for this sample) over
the area of the WFPC1 field of view predicts an inner to outer ratio of~1.49
consistent with the directly measured star count ratio.  On the other hand,
integration of the $22''$ profile that best fits the star list and PSF-fitting
photometry used in this paper (case HHH in Table~\ref{table1}) yields
an inner to outer ratio of~1.05; this is comparable to the directly measured
star count ratio of $1.02\pm0.03$.

\section{Conclusions}

This paper presents estimates of the density profile of the globular cluster
47~Tuc based on three samples of stars (star list and photometry) derived
independently from {\it Hubble Space Telescope\/} WFPC1 and WFPC2 images.
Apparent discrepancies amongst the core radius measurements published by
\cite{gysb}, \cite{guido}, and \cite{calzetti} are investigated.
Our conclusion is that there is severe, radially-varying bias in the
magnitude-limited star counts used by De~Marchi et~al.\ and Calzetti et~al.,
and this causes their core radius estimates to be spuriously low ($r_{\rm
core}\sim14''$) relative to other determinations ($r_{\rm core}\sim23''$).
This ``limiting magnitude bias'' is a result of
large photometric scatter/bias associated with the application of their
aperture photometry method to the crowded central regions of the cluster,
coupled with a choice of limiting magnitude near the steep part of the
stellar luminosity function.  In general, such a choice of limiting magnitude
is dangerous; even with symmetric errors the resulting sample will be
contaminated by large numbers of stars just fainter than the cutoff.
Any radial variation in the magnitude of the errors will cause the degree
of this contamination to vary, resulting in an incorrect determination of the
radial density profile.

Combining De~Marchi et~al.'s photometry with a limiting magnitude near the
main sequence turnoff at the peak of the luminosity function reduces, but
does not eliminate, the discrepancy ($r_{\rm core}\sim18''$); the radial
variations in DPSGB's photometry (larger photometric scatter/bias at small
radii) have a significant effect on the derived core radius even with an
optimal choice of limiting magnitude.
A more accurate PSF-fitting method is used in this paper to indepedently
derive two sets of stellar photometry, one from the deep WFPC1 data
analyzed by De~Marchi et~al.\ and the other from an archival WFPC2 data set.
The core radii derived using these two photometry sets are independent of the
choice of limiting magnitude and star list, and are consistent with each
other and with previous ground-based and {\it HST\/} work: $r_{\rm
core}\sim23''$ (cf.~\cite{hr79}; \cite{dk84}; Guhathakurta et~al.\ 1992).

The best fit core radius for the surface density distribution of
evolved stars in 47~Tuc is about 15\% of the cluster half-mass radius
($r_{\rm h}=174''$).  This is significantly larger than the range of $(r_{\rm
core}/r_{\rm h})$ values found in numerical simulations of
post--core-collapse clusters: 0.01--0.04 (\cite{cohn80}; \cite{good87};
\cite{gao91}).  It should be noted however that the surface brightness
profile is not a perfect discriminant between a relaxed and
post--core-collapse cluster, and it is advisable to combine it with velocity
dispersion data (Gebhardt \& Fischer 1995).

\bigskip
\bigskip
\acknowledgments

We are grateful to Guido De~Marchi for making stellar photometry tables
available to us in electronic form, and to Randi Cohen for help in the early
phase of this project.  We would like to thank Fernando Camilo, Paulo Freire,
Karl Gebhardt, and Fred Rasio for helpful discussions, and the referees,
especially Tad Pryor, for several useful suggestions.

\end{document}